\documentclass[aps,prl,preprint,unsortedaddress,showkeys,nofootinbib]{revtex4-1}

\usepackage{amssymb,amsmath,amsfonts}

\usepackage{graphics}
\usepackage{graphicx}
\usepackage{epstopdf}
\usepackage[pdftex]{hyperref}
\usepackage{longtable}

\usepackage{bm}
\usepackage{subfigure}

\usepackage{color}
\usepackage{setspace}
\usepackage{multirow}

\usepackage[ruled,vlined,linesnumbered]{algorithm2e}

\usepackage{color, colortbl}
\definecolor{Yellow}{rgb}{1,1,0}
\definecolor{Grey}{rgb}{.87,.87,.87}
\definecolor{Purple}{rgb}{.8,.0,1.0}
\definecolor{Crimson}{rgb}{.86,.08,.23}

\begin{document}

\title{Recurrent Graph Neural Network Algorithm for Unsupervised Network Community Detection}

\author{Stanislav Sobolevsky
\footnote{To whom correspondence should be
addressed: sobolevsky@nyu.edu}}
\affiliation{
Center For Urban Science+Progress, New York University, Brooklyn, NY, USA\\
}

%\contributor{}

%\begin{article}

\date{\today}

\begin{abstract}
\begin{it}
Network community detection often relies on optimizing partition quality functions, like modularity. This optimization appears to be a complex problem traditionally relying on discrete heuristics. And although the problem could be reformulated as continuous optimization, direct application of the standard optimization methods has limited efficiency in overcoming the numerous local extrema. However, the rise of deep learning and its applications to graphs offers new opportunities. And while graph neural networks have been used for supervised and unsupervised learning on networks, their application to modularity optimization has not been explored yet. This paper proposes a new variant of the recurrent graph neural network algorithm for unsupervised network community detection through modularity optimization. The new algorithm's performance is compared against a popular and fast Louvain method and a more efficient but slower Combo algorithm recently proposed by the author. The approach also serves as a proof-of-concept for the broader application of recurrent graph neural networks to unsupervised network optimization.
\end{it}
\end{abstract}


\maketitle

\section*{Introduction}
The complex networks play a pivotal role in various fields such as physics, biology, economics, social sciences, urban planning. Thus understanding the underlying community structure of the networks saw a wide range of applications, including social science \cite{plantie2013survey}, biology\cite{Guimera2005FunctionalCartography}, economics \cite{PiccardiWorldTradeWeb}. In particular, partitioning the networks of human mobility and interactions is broadly applied to regional delineation \cite{Ratti2010GB, blondel2010regions, Sobolevsky2013delineating, amini2014impact, hawelka2014geo, kang2013exploring, sobolevsky2014money, belyi2017global, grauwin2017identifying} as well as urban zoning \cite{sobolevsky2018twitter, landsman2020zoning, landsman2021social}.

Over the last two decades a big number of approaches and algorithms for community detection in complex networks has been suggested. Some of them are just the straightforward heuristics such as hierarchical clustering\,\cite{Hastie2001ElementsOfStatisticalLearning} or the Girvan-Newman\,\cite{GN} algorithm, while the vast majority rely on optimization techniques based on the maximization of various objective functions. The first and the most well-known partition quality function is modularity\,\cite{newman2004,newman2006} assessing the relative strength of edges and quantifying the cumulative strength of the intra-community links. A large number of modularity optimization strategies have been suggested over the last two decades \cite{NewmanPRE2004, CNM2004VeryLargeNetworks, newman2004, newman2006, Sun2009, louvain, simulatedAnnealing,Good2010PerformanceOfModularity, Duch2005CElegans, LeeCSA, combo}. A comprehensive historical overviews are presented in \cite{fortunato2010, fortunato2016community} as well as some later surveys \cite{khan2017network, javed2018community}.

And while the problem of finding exact modularity maximum is known to be NP-hard \cite{brandes2006maximizing}, most of the available modularity optimization approaches rely on certain discrete optimization heuristics (although, in some cases an algorithmic optimality proof of the partition is possible \cite{sobolevsky2017optimality}).

As we show below the modularity optimization can be formulated as a continuous matrix optimization problem, however direct application of generic gradient descent methods is not efficient due to a large number of local maxima, that gradient descent might not be able to overcome.

Since recently graph neural networks (GNNs) have became increasingly popular for supervised classifications and unsupervised embedding of the graph nodes with diverse applications in text classification, recommendation system, traffic prediction, computer vision etc  \cite{wu2020comprehensive}. And GNNs were already successfully applied for community detection, including supervised learning of the ground-truth community structure \cite{chen2017supervised} as well as unsupervised learning of the node features enabling representation modeling of the network, including stochastic block-model \cite{bruna2017community} and other probabilistic models with overlapping communities \cite{shchur2019overlapping} or more complex self-expressive representation  \cite{bandyopadhyay2020self}. However, existing GNN applications overlook unsupervised modularity optimization, which so far has been a major approach in classic community detection.

This paper aims to fill this gap by proposing a straightforward GNN-inspired algorithmic framework for unsupervised community detection through modularity optimization. 
We perform a comprehensive comparative evaluation of the performance of the proposed method against state-of-the-art Combo algorithm (capable of reaching the best known partition in most cases) and a viral Louvain algorithm (which, despite its sub-optimal performance, is often comparatively fast and capable of handling large-scale networks). We demonstrate that the method provides a reasonable balance between performance and speed for classic, synthetic and real-world networks, including temporal networks, and is sometimes capable of finding partitions with a higher modularity score that other algorithms cannot achieve. More importantly, we believe the proposed approach serves as a proof-of-concept of leveraging GNN approaches for solving a broader range of network optimization problems.

\section{The modularity optimization problem}

The network modularity was among the first quality/objective functions proposed to assess and optimize the community structure \cite{newman2006}. It is now known to have certain shortcomings including a resolution limit \cite{Fortunato02012007ResolutionLimit, Good2010PerformanceOfModularity}. Therefore alternative objective functions deserve to be mentioned, e.g. Infomap description code length \cite{Rosvall01052007InformationTheoretic, Infomap}, Stochastic Block Model likelihood \cite{Newman2011Stochastic,Newman2011Efficient,Bickel2009Nonparametric, Decelle2011BlockModel, Decelle2011BlockModelAsymptotics, Yan2012ModelSelection}, and Surprise \cite{Aldecoa2011Deciphering}. However despite its limitations, modularity remains to be perhaps the most commonly used objective function so far. 

Recently the authors proposed a novel optimization technique for community detection "Combo" \cite{combo} capable of maximizing various objective functions, including modularity, description code length and pretty much any other metric based on the link scoring and assessing the cumulative score of the intra-community links. For modularity optimization Combo outperforms other state-of-the-art algorithms, including a popular Louvain method \cite{louvain} in terms of the quality (modularity score) of the resulting partitioning which could be achieved within a reasonable time for the most of the real-world and synthetic networks of up to tens of thousands of nodes. The size limitation for the algorithm evaluation is due to the current implementation handling a full modularity matrix in the memory entirely. However this is not a fundamental limitation and it could be overcome by using sparse matrix operations.

The proposed algorithms, including Combo are often quite efficient and in some cases, are able to reach the theoretic maximum of the modularity score as revealed by a suitable upper bound estimate \cite{sobolevsky2017optimality}. However in general finding the theoretically optimal solution may not be feasible and one has to rely on heuristic algorithmic solutions, without being certain of their optimality. Instead, empiric assessment of their performance in comparison with other available algorithms could be performed.

\subsection{The modularity function}
In short, the modularity ~\cite{newman2004,newman2006} function of the proposed network partition quantifies how relatively strong are all the edges between the nodes attached to the same community.
Specifically, if the weights of the network edges between each pair of nodes $i,j$ are denoted as $e_{i,j}$, then the modularity of the partition $com(i)$ (expressed as a mapping assigning community number $com$ to each node $i$) can be defined as 
\begin{equation}
M=\sum_{i,j, com(i)=com(j)}q_{i,j},
\label{modularity}
\end{equation}
where the quantity $q_{i,j}$ for each edge $i,j$ (call $q$ a modularity score for an edge) is defined as its normalized relative edge weight in comparison with the random network model with the same node weights. Namely,
$$
q_{i,j}=\frac{e_{i,j}}{T}-\frac{w^{out}(i)w^{in}(j)}{T^2},
$$
where $w^{out}(i)=\sum_k e_{i,k}$, $w^{in}(j)=\sum_k e_{k,j}$, $T=\sum_i w^{out}(i)=\sum_j w^{in}(j)=\sum_{i,j}e_{i,j}$.

Rewrite the modularity optimization problem in a vector form. Let $Q=\left(q_{i,j}\right)$ be the matrix of all the modularity scores for all the edges (call it a modularity matrix).

Let $C$ be an $n\times m$ matrix, where $n$ is the number of network nodes and $m$ is the number of communities we're looking to build. Each element $c_{i,p}$ of the matrix can be zero or one depending on whether the node $i$ belongs to the community $p$ or not, i.e. whether $com(i)=p$. If the communities are not overlapping then each row of the matrix has one single unit element and the rest of its elements are zeros.

More generally, if we admit uncertainty in community attachment, the elements $c_{i,p}$ of the matrix $C$ could then represent the probabilities of the node $i$ to be attached to the community $p$. This way, $c_{i,p}\in [0,1]$ and the sum of each row of the matrix $C$ equals to $1$.

Then the modularity score $M$ in case of a discrete community attachment could be represented as a matrix product
\begin{equation}
M=tr(C^T Q C),
\label{modularity_matrix}
\end{equation}
where $tr$ denotes the trace of the matrix - a sum of all of its diagonal elements.

This way, finding the community structure of up to $m$ communities optimizing the network modularity could be expressed as a constrained quadratic optimization problem of finding the $n\times m$ matrix $C$ maximizing the trace of the matrix product $M=C^T Q C$, such that all $c_{i,p}\in \{0,1\}$, and the sum of each row of the matrix $C$ equals to $1$ (having a single unit element). 

Replacing the binary attachment constraint $c_{i,p}\in \{0,1\}$ with a continuous attachment $c_{i,p}\in [0,1]$ relaxes the optimization problem to finding probabilistic community attachments. It could be easily shown that the optimal solution of the binary attachment problem could be derived from the optimal solution of the probabilistic attachment problem after assigning $q_{i,i}=0$ for all the diagonal elements of the matrix $Q$. As diagonal elements $q_{i,i}$ are always included into the sum $M$ since $com(i)=com(j)$ for $i=j$, the values of the diagonal elements serve as constant adjustment of the objective function $M$ and do not affect the choice of the optimal partition, so we are free to null them without loss of generality. At the same time, for each given $i$, once $q_{i,i}=0$, if we fix the community attachments of all the other nodes $j\neq i$, the objective function $M$ becomes a linear function of the variables $c_{i,p}$ subject to constraints $\sum_p c_{i,p}=1$ and $c_{i,p}\in \{0,1\}$. Obviously, the maximum of the linear function with linear constraints is reached at one of the vertices of the domain of the allowed values for $c_{i,p}$, which will involve a single $c_{i,p}$ being one and the rest being zeros. This way, we've proven the following: 

{\bf Proposition.} The optimal probabilistic attachment $c_{i,p}\in [0,1]$ maximizing (\ref{modularity_matrix}) in case of $q_{i,i}=0$ represents a binary attachment $c_{i,p}\in \{0,1\}$ maximizing (\ref{modularity_matrix}) for an arbitrary original $Q$.

So the discrete community detection problem through modularity optimization could be solved within the continuous constrained quadratic optimization framework. However, despite its analytic simplicity, the dimensionality of the problem leads to multiple local maxima challenging direct application of the standard continuous optimization techniques, like gradient descent. Indeed, any discrete partition, such that no single node could be moved to a different community with a modularity gain, will become such a local maxima. Unfortunately, finding such a local maxima rarely provides a plausible partition - such solutions could have been obtained with a simple greedy discrete heuristic iteratively adjusting the single node attachments, while we know that the modularity optimization, being NP-hard, generally requires more sophisticated non-greedy heuristics, like \cite{combo}.

\section{The GNNS method}

This paper introduces a GNN-style method (GNNS) for unsupervised network partition through modularity optimization inspired by recurrent graph neural network models (in definition of \cite{wu2020comprehensive}) as well as an older Weisfeiler-Lehmann graph node labeling algorithm \cite{weisfeiler1968reduction}. Namely, implement a simple iterative process starting with a random initial matrix $C=C_0$ and at each step $k=1,2,3,...,N$ performing an iterative update of the rows $c_i$ of the matrix $C$ representing the node $i$ community attachments as following:
\begin{equation}
\tilde{c}_{i}^k = F\left(c_i^{k-1}, Q_i C^{k-1}\right),
\label{recGNN}
\end{equation}
where $Q_i=\left(q_{i,j}:j=1,2,...,n\right)$ is the $i$-th row of the modularity matrix $Q$ representing the outgoing edges from the node $i$. This way the term $Q_i C^{k-1}$ collects information about the neighbor nodes' community attachments and the equation (\ref{recGNN}) updates the node community attachments with respect to their previous attachments as well as the neighbor node attachments.
In order to ensure the conditions $\sum_p c_{i,p}=1$, a further normalization $c_{i,p}^k = \tilde{c}_{i,p}^k/\sum_{p^*}\tilde{c}_{i,p^*}^k$ needs to be applied at each iteration.

A simple form for an activation function $F$ could be a superposition of a linear function subject to appropriate scale normalization and a %two-side variant of a 
%$ReLU(x)=\left\{\begin{array}{c}0,x<0\\x,x\in[0,1]\\1,x>1\end{array}\right.$, i.e.
rectified linear unit 
$ReLU(x)=\left\{\begin{array}{c}0,x\leq 0\\x,x>0\end{array}\right.$, leading to
\begin{equation}
\tilde{c}_{i,p}^k=ReLU\left(f_1 c_{i,p} + f_2 Q_i C_p^{k-1}/t_i^k + f_0\right), \ c_{i,p}^k = \tilde{c}_{i,p}^k/\sum_{p^*}\tilde{c}_{i,p^*}^k,
\label{recGNN1}
\end{equation}
where $f_0, f_1, f_2$ are the model parameters, $C_p=\left(c_{j,p}:j=1,2,...,n\right)$ is the $p$-th column of the matrix $C$ representing all the node attachments to the community $p$, and the $t_i^k = \max_{p^*} Q_i C_{p^*}^{k-1}$ are the normalization coefficients ensuring the same scale for terms of the formulae.

Intuitive considerations allow defining possible ranges for the model coefficients $f_0,f_1,f_2$.
Defining the coefficient $f_1$ within the range $f_1\in [0,1]$ would ensure decay scaling of the community attachment at each iteration unless confirmed by the strength of the node's attachment to the rest of the community expressed by $Q_i C_p^{k-1}$. A free term $f_0\in [-1,0]$ provides some additional constant decay of the community attachment at each iteration. While the term $Q_i C_p^{k-1}/t_i^k$ strengthens attachment of the node $i$ to those communities having positive modularity scores of the edges between the node $i$ and the rest of the community. Normalization term $t_i^k$ ensures that the strongest community attachment gets a maximum improvement of a fixed scale $f_2$.

A consistent node attachment that can't be improved by assigning the give node $i$ to a different community $p$ (i.e. having $Q_i C_p=t_i=\max_{p^*} Q_i C_{p^*}^{k-1}$), should see a fastest increase in the attachment score $\tilde{c}_{i,p}$ eventually converging to the case of $c_{i,p}=1$. Any weaker attachment should see a decreasing community membership score $c_{i,p}$ eventually dropping to zero. This could be ensured by the balancing equation $f_1+f_2+f_0 = 1$, allowing to define appropriate $f_2$ given $f_0$ and $f_1$.

\section{Training the GNNS}

The sequence of the GNNS iterations (\ref{recGNN1}) depends on the choice of the model parameters $f_0,f_1$ as well as the initial community attachments. And the final convergence might often depend on those choices. Given that a good strategy is to simulate multiple iteration sequences with different initial attachments and choose the best final result. Also it turns out that the method demonstrates reasonable performance for a broad range of parameter values $f_0\in [-1,0]$, $f_1\in [0,1]$, so rather than trying to fit the best choice for all the network or for a given network, one may simply include the random choice for $f_0, f_1$ along with a random choice of the initial community attachments.

So the proposed GNNS algorithm starts with a certain number of $S$ random partitions and parameter choices. Then the GNNS performs 10 iterations of the partition updates according to (\ref{recGNN1}). Among those a batch of the best $[S/3]$ partitions (with highest achieved modularity scores) is selected and further supplemented with another $S-[S/3]$ configurations derived from the selected batch by randomly shuffling partitions and parameters (i.e. picking the values of $f_0, f_1$ from one and the achieved partition from another random member of the selected batch). Another 10 iterations are performed. Another batch of $[S/9]$ best partitions is selected and shuffled creating a total of $[S/3]$ samples. Then the final 30 iterations are performed with those and the best resulting partition is selected as the final outcome of the algorithm.

Below we evaluate the two versions of the GNNS: a fast version with $S=100$ denoted GNNS100 and a slower but more precise version with $S=2500$ denoted GNNS2500.

\section{Comparative evaluation}

We evaluate a proposed GNNS modularity optimization algorithm against the two state-of-the-art techniques mentioned above - a fast and popular Louvain method \cite{louvain} and a Combo algorithm \cite{combo} often capable of reaching the best known modularity score over a wide range of networks. We shall compare the three approaches over a sample of classic network examples as well as the series of the two types of random graphs often used for benchmarking the community detection algorithms - Lancichinetti-Fortunato-Radicchi \cite{LFR} and Block-model graphs \cite{Newman2011Stochastic}.

As the GNNS chooses the best partition among multiple runs we consider its two configurations involving a) 100 and b) 2500 of initial random samples of partitions+model configurations. We shall refer to those as GNNS100 and GNNS2500. The Combo and Louvain algorithms also involve random steps and the best partition they converge to is not perfectly stable. Thus their performance could also benefit from choosing the best partition among multiple runs, especially for the Louvain method. It often takes up to 10-20 attempts to find the best partition they are capable of producing. E.g. applying them to the classic case of the Email leads to the following performance reported in the table \ref{tab:email_performance} below. As we see both reach their best performance after 20 iterations (although results of the Combo for this network are significantly better compared to Louvain) but do not further improve over the next 30 iterations. Based on that in the further experiments we shall report the best performance of the Combo and Louvain algorithms achieved after 20 attempts for each.

\begin{table*}[htpb]
\caption{\label{tab:email_performance}The best modularity score reached by the Combo and Louvain method after different number of attempts for the Email network.}

\begin{center}
\begin{tabular}{| l | c | c | c | c | c |}
\hline
Method/attempts& 1 & 5 & 10 & 20 & 50 \\
\hline
Combo & 0.581918 & 0.582751 & 0.582751 & 0.582829 & 0.582829\\ 
Louvain & 0.563761 & 0.570319 & 0.573820 & 0.574912 & 0.574912\\
\hline
\end{tabular}
\\[10pt]
\end{center}
\end{table*}

\subsection{Classic examples}
Consider the performance of the proposed approach on several classical examples of networks (the sources and details of those networks are introduced in the appendix in the table \ref{tab:networksList}) in comparison with Louvain \cite{louvain} and Combo methods \cite{combo}. According to the results reported in the table \ref{tab:networkScores}, the GNNS100 is the fastest algorithm on average trailed by Louvain method, while the GNNS2500 is by far the slowest, however demonstrates superior average performance. Its average advantage over Combo is minimal, besides Combo reaches the best performance on a larger number of networks (9 vs 8 out of 12). And given that it is an order of magnitude faster, Combo looks like the most balanced solution when the partition quality is prioritized. However GNNS2500 was capable of noticeably improving the best Combo result for 3 out of 12 networks and for the Copperfield example GNNS2500 achieved the modularity score way above those reached by any of the other algorithms, which justifies its application along with other methods when reaching the best possible score is of the paramount importance. On the other side of the spectrum, the GNNS100 is considerably faster than all the other methods including Louvain and several times faster than Combo for larger networks. And while GNNS100 demonstrates reasonable overall performance comparable with other methods and better on average than Louvain, the GNNS100 could be recommended as the most balanced choice when the priority is the computational speed.

Besides, the current GNNS implementation uses pure Python, while moving it to C++ (as done for Combo) could provide further speed improvement. The additional code optimization is also possible through replacing full matrix multiplication by sparse matrix operations, that could noticeably improve the performance for larger sparse networks. So the full benefits of applying GNNS to large sparse networks are yet to be explored.

Overall, while no single heuristic is the best solution for all the cases, GNNS algorithm has its niche while applied to classical networks.

\begin{table*}[htpb]
\footnotesize
\caption{\label{tab:networkScores}Comparative evaluation of the Louvain, Combo and GNNS algorithms over the classical network examples. The sources of the networks are detailed in the appendix. Those networks marked with asterisk were symmetrized before partition in order to enable application of all the three methods.}

\begin{center}
\begin{tabular}{| c | p{2.5cm} | p{1.6cm} | p{1.6cm} | p{1.6cm} | p{1.6cm} | p{1.6cm} | p{1.3cm} | p{1.3cm} | p{1.3cm} | p{1.3cm} | p{1.3cm} | p{1.3cm}}
\hline
\textbf{No} & \textbf{Name} & \multicolumn{4}{c|}{Best modularity score} & \multicolumn{4}{c|}{Time, sec}\\
\cline{3-10}
& & Louvain & Combo & GNNS 100 & GNNS 2500 & Louvain & Combo & GNNS 100 & GNNS 2500\\
\hline
1 & Karate & \bf 0.419790 & \bf 0.419790 & \bf 0.419790 & \bf 0.419790 & 0.06 & \bf 0.02 & 0.06 & 3.19\\
2 & Dolphins & \bf 0.528519 & 0.526799 & \bf 0.528519 & \bf 0.528519 & 0.12 & \bf 0.05 & 0.07 & 2.04\\
3 & Les Miserable & \bf 0.566688 & \bf 0.566688 & \bf 0.566688 & \bf 0.566688 & 0.12 & \bf 0.09 & \bf 0.09 & 5.04\\
4 & Football & \bf 0.605445 & \bf 0.605445 & 0.604274 & \bf 0.605445 & 0.22 & \bf 0.20 & 0.22 & 7.03\\
5 & Political Books & \bf 0.527237 & \bf 0.527237 & \bf 0.527237 & \bf 0.527237 & 0.24 & 0.18 & \bf 0.14 & 3.88\\
6 & Copperfield & 0.306007 & 0.310580 & 0.309229 & \bf 0.313359 & 0.29 & 0.26 & \bf 0.16 & 4.74\\
7 & Jazz$^*$ & \bf 0.445144 & 0.444871 & \bf 0.445144 & \bf 0.445144 & 0.90 & 0.36 & \bf 0.31 & 8.98\\
8 & C. Elegans$^*$ & 0.499271 & \bf 0.503782 & 0.503483 & \bf 0.503782 & 1.02 & 1.22 & \bf 0.45 & 12.23\\
9 & Airports97$^*$ & 0.204002 & \bf 0.214688 & 0.210993 & 0.214632 & 1.18 & 1.89 & \bf 0.51 & 16.04 \\
%10 & Metabolic$^*$ & 0.445239 & 0.451291 & \bf 0.451320 & 0.445119 & 0.447276 & 1.09 & 5.42 & 56.42 & \bf 1.05 & 36.62\\
10 & Emails & 0.574912 & \bf 0.582739 & 0.567555 & 0.576774 & \bf 4.49 & 46.93 & 4.97 & 167.58 \\
11 & Blogs$^*$ & 0.432248 & \bf 0.432465 & 0.432222 & 0.432433 & 9.52 & 29.67 & \bf 8.45 & 213.61 \\
12 & Airports2010$^*$ & 0.274288 & \bf 0.275524 & 0.274659 & 0.275502 & \bf 10.60 & 60.54 & 10.82 & 332.66 \\
\hline
 & Avg \% to best & 99.162 & 99.894 & 99.478 & 99.911 & 183.086 & 284.697 & 121.897 & 4419.759 \\
\hline

\end{tabular}
\\[10pt]
\end{center}
\end{table*}

\subsection{Synthetic networks}

In the next test, the methods were applied to the sets of synthetic networks - Lancichinetti–Fortunato–Radicchi (LFR) \cite{LFR} and Stochastic Block-Model (SBM) \cite{holland1983stochastic}. We build 10 LFR networks of size 250 each and 10 SBM networks of size 300 for each value of the parameter $\nu$, defining the ratio of the probability for the model graph to have inner-community edges (three communities of size 100 each) divided by the probability of the inter-community edges. The Python package networkx was used to generate these synthetic networks.

The average results of partitioning the 10 models of each type using Louvain, Combo (best of the 20 runs) and GNNS100 algorithms are presented in the table \ref{tab:synethetic} below. As we can see all three algorithms give the same average modularity and average normalized mutual information (NMI) between the resulting and original model communities for LFR and SBM with the highest $\nu=3$. While for the $\nu=1.5,2.0,2.5$ the Combo demonstrates superior performance trailed by GNNS100, in turn, substantially outperforming the Louvain method. And although GNNS100 demonstrates suboptimal performance for those models compared to Combo, it works at unparalleled speed several times faster than Combo and Louvain. Based on that, GNNS100 proves itself to be by far the fastest solution for partitioning the synthetic networks demonstrating reasonable performance in terms of the modularity and NMI scores achieved.

\begin{table*}[htpb]
\footnotesize
\caption{\label{tab:synethetic}Comparative evaluation of the Louvain, Combo and GNNS algorithms over the synthetic networks.}
\begin{center}
\begin{tabular}{| p{2cm} | p{1.5cm} | p{1.5cm} | p{1.5cm} | p{1.5cm} | p{1.5cm} | p{1.5cm} | p{1.3cm} | p{1.3cm} | p{1.3cm} |} 
\hline
\textbf{Model} & \multicolumn{3}{c|}{Avg. modularity score} & \multicolumn{3}{c|}{Avg. NMI} & \multicolumn{3}{c|}{Avg. time, sec}\\
\cline{2-10}
& Louvain & Combo & GNNS 100 & Louvain & Combo & GNNS 100 & Louvain & Combo & GNNS 100\\
\hline
LFR & 0.665854 & 0.665854 & 0.665854 & 0.545034 & 0.545034 & 0.545034 & 1.054 & 1.059 & 0.565\\
SBM $\nu=1.5$ & 0.157852 & 0.170889 & 0.164369 & 0.036434 & 0.036967 & 0.052729 & 2.963 & 3.222 & 0.601\\
SBM $\nu=2.0$ & 0.167859 & 0.183892 & 0.177800 & 0.204417 & 0.417851 & 0.310943 & 3.135 & 3.221 & 0.584\\
SBM $\nu=2.5$ & 0.221330 & 0.222546 & 0.222463 & 0.852960 & 0.892313 & 0.881727 & 3.363 & 1.564 & 0.369\\
SBM $\nu=3.0$ & 0.267409 & 0.267409 & 0.267409 & 0.963604 & 0.963604 & 0.963604 & 2.684 & 1.269 & 0.502\\
\hline
\end{tabular}
\\[10pt]
\end{center}
\end{table*}

\subsection{Temporal networks}

Earlier works established applicability of the GNN architecture for capturing dynamic properties of the evolving graphs \cite{ma2020streaming}. As GNNS is well-suited for the iterative partition improvement, it could be suggested for active learning of the temporal network partition. Initial warm-up training could be performed over the first temporal layers with subsequent tuning iterations while moving from a current temporal layer to the next one.

Below we apply the approach to the temporal network of the daily taxi mobility between the taxi zones in the New York City. We use the 2016-2017 data provided by the NYC Taxi and Limousine Commission \footnote{https://www1.nyc.gov/site/tlc/about/data.page} to build the origin-destination network of yellow and green taxi ridership between the NYC taxi zones (edges of the network are weighted by the number of trips). The results of the temporal GNNS (tempGNNS) for each daily ridership network are compared against single runs (for the sake of speed) of Combo and Louvain algorithms. The tempGNNS uses an initial warm up over the year 2016 aggregated network and then performs a single run of 20 fine-tune iterations for each daily temporal layer in 2017 starting with the previously achieved partition. The achieved best modularity scores fluctuate slightly between the daily layers. The 2017 yearly average of the ratios of the daily scores achieved by each algorithm to the best score of all three algorithms for that day look as following:
$99.02\%$ for Louvain, $99.92\%$ for Combo and $99.50\%$ for tempGNNS. While the total elapsed time is as following: 83.35 sec for Louvain, 50.97 sec for Combo and 27.89 sec for tempGNNS. Furthermore, tempGNNS managed to find the best modularity score not reached by the two other networks on $4.1\%$ of the temporal layers. 

So performance of the tempGNNS in terms of the achieved modularity score falls right in the middle between Louvain and Combo, while the tempGNNS is nearly twice as fast as the single runs of Combo and more than three times faster than Louvain. 

\begin{figure}
    \centering
    \includegraphics[width=16.5cm]{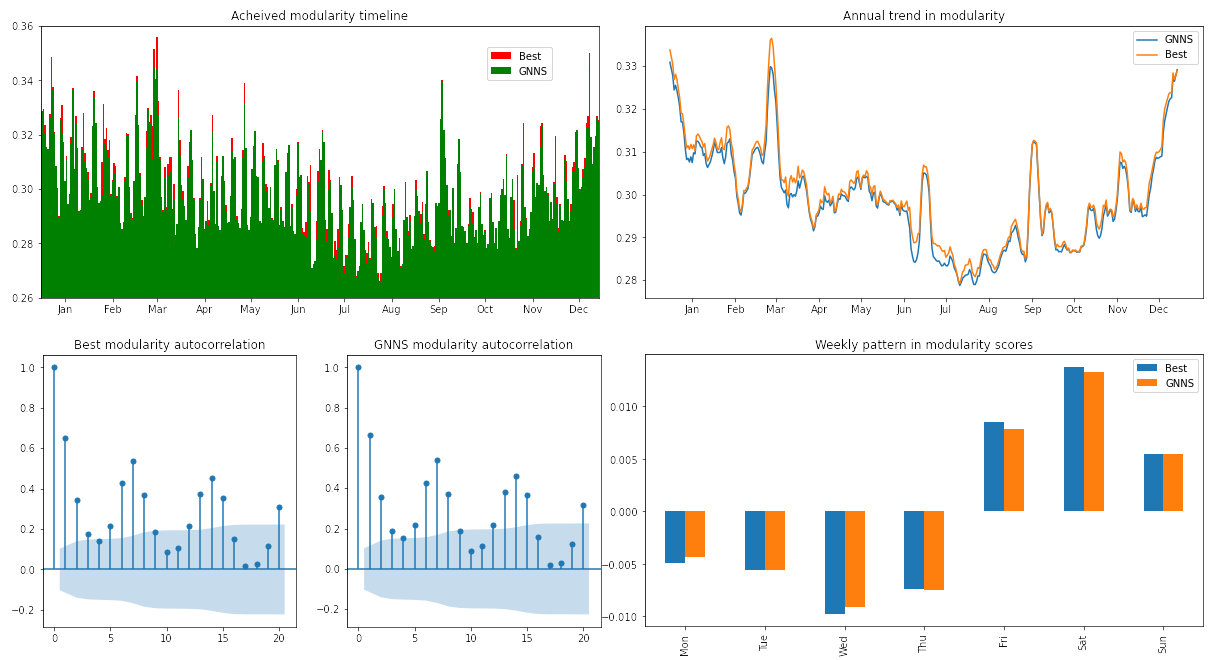}
    \caption{Comparative performance of the GNNS vs the best of GNNS, Louvain and Combo for community detection on the temporal network of daily taxi ridership in NYC. Achieved network modularity and its time-series properties: seasonality and periodicity.}
    \label{fig:GNNStemp}
\end{figure}

Details of the day-by-day tempGNNS performance compared to the best of the three algorithms are shown on the figure \ref{fig:GNNStemp} below. 
On some days one may notice somewhat visible differences between the top algorithm performance and that of the tempGNNS, however tempGNNS captures the temporal pattern of the modularity score dynamics pretty well with the correlation of $99.02\%$ between the tempGNNS score and the top algorithm score timelines.

By performing time-series periodicity and trend-seasonality analysis, one could notice some interesting temporal patterns in the strength of the network community structure as also presented on the figure \ref{fig:GNNStemp}. The community structure quantified by means of the best achieved modularity score demonstrates a strong weekly periodicity with the stronger community structure over the weekends, including Fridays. The strength of the community structure also shows noticeable seasonality with stronger communities over the winter time and weaker over the summer. One may relate this observation with the people exploring more destinations during their weekend and holiday time. As seen on the figure \ref{fig:GNNStemp} the tempGNNS accurately reproduces the patterns discovered for the best partition timeline.

In conclusion, while Combo runs could demonstrate higher partition accuracy at certain temporal layers, the tempGNNS is capable of reaching reasonable performance at unparalleled speed, while adequately reproducing the qualitative temporal patterns. So tempGNNS could be trusted as a fast and efficient solution for extracting insights on the dynamics of the temporal network structure.

\section{Conclusions}
We proposed a novel recurrent GNN-inspired framework for unsupervised learning of the network community structure through modularity optimization. A simple iterative algorithm depends on only two variable parameters, and we propose an integrated technique for tuning those so that parameter selection and modularity optimization are performed within the same iterative learning process. 

The algorithm's performance has been evaluated on the classic network examples, synthetic network models, and a real-world network case. Despite its simplicity, the new algorithm noticeable outperforms a popular Louvain algorithm. It reaches comparable and, in some cases, higher modularity scores compared to the more sophisticated discrete optimization algorithm Combo, which so far has consistently demonstrated superior performance in comparison with other state-of-the-art algorithms. Furthermore, the algorithm allows flexible adjustment of its complexity in tuning the model parameters to find the right balance between speed and performance. At the low-complexity settings, it can significantly outperform alternative methods in terms of running time while maintaining reasonable performance in terms of partition quality. Thus, the algorithm is efficiently applicable in both scenarios - when the execution time is of the essence as well as when the quality of the resulting partition is a paramount priority. 

Furthermore, the algorithm enables a special configuration for the active learning of the community structure on temporal networks, reconstructing all the important longitudinal patterns at unparalleled speed.

But more importantly, we believe the algorithm serves as a successful proof of concept for applying GNN-inspired techniques for unsupervised network learning and opens possibilities for solving a broader range of network optimization problems.

\bibliographystyle{apsrev}
\bibliography{ModularityGNN}

\begin{thebibliography}{62}
\expandafter\ifx\csname natexlab\endcsname\relax\def\natexlab#1{#1}\fi
\expandafter\ifx\csname bibnamefont\endcsname\relax
  \def\bibnamefont#1{#1}\fi
\expandafter\ifx\csname bibfnamefont\endcsname\relax
  \def\bibfnamefont#1{#1}\fi
\expandafter\ifx\csname citenamefont\endcsname\relax
  \def\citenamefont#1{#1}\fi
\expandafter\ifx\csname url\endcsname\relax
  \def\url#1{\texttt{#1}}\fi
\expandafter\ifx\csname urlprefix\endcsname\relax\def\urlprefix{URL }\fi
\providecommand{\bibinfo}[2]{#2}
\providecommand{\eprint}[2][]{\url{#2}}

\bibitem[{\citenamefont{Planti{\'e} and Crampes}(2013)}]{plantie2013survey}
\bibinfo{author}{\bibfnamefont{M.}~\bibnamefont{Planti{\'e}}} \bibnamefont{and}
  \bibinfo{author}{\bibfnamefont{M.}~\bibnamefont{Crampes}}, in
  \emph{\bibinfo{booktitle}{Social media retrieval}}
  (\bibinfo{publisher}{Springer}, \bibinfo{year}{2013}), pp.
  \bibinfo{pages}{65--85}.

\bibitem[{\citenamefont{Guimer\`a and
  Nunes~Amaral}(2005)}]{Guimera2005FunctionalCartography}
\bibinfo{author}{\bibfnamefont{R.}~\bibnamefont{Guimer\`a}} \bibnamefont{and}
  \bibinfo{author}{\bibfnamefont{L.~A.} \bibnamefont{Nunes~Amaral}},
  \bibinfo{journal}{Nature} \textbf{\bibinfo{volume}{433}},
  \bibinfo{pages}{895} (\bibinfo{year}{2005}), ISSN \bibinfo{issn}{0028-0836},
  \urlprefix\url{http://dx.doi.org/10.1038/nature03288}.

\bibitem[{\citenamefont{Piccardi and Tajoli}(2012)}]{PiccardiWorldTradeWeb}
\bibinfo{author}{\bibfnamefont{C.}~\bibnamefont{Piccardi}} \bibnamefont{and}
  \bibinfo{author}{\bibfnamefont{L.}~\bibnamefont{Tajoli}},
  \bibinfo{journal}{Phys. Rev. E} \textbf{\bibinfo{volume}{85}},
  \bibinfo{pages}{066119} (\bibinfo{year}{2012}),
  \urlprefix\url{http://link.aps.org/doi/10.1103/PhysRevE.85.066119}.

\bibitem[{\citenamefont{Ratti et~al.}(2010)\citenamefont{Ratti, Sobolevsky,
  Calabrese, Andris, Reades, Martino, Claxton, and Strogatz}}]{Ratti2010GB}
\bibinfo{author}{\bibfnamefont{C.}~\bibnamefont{Ratti}},
  \bibinfo{author}{\bibfnamefont{S.}~\bibnamefont{Sobolevsky}},
  \bibinfo{author}{\bibfnamefont{F.}~\bibnamefont{Calabrese}},
  \bibinfo{author}{\bibfnamefont{C.}~\bibnamefont{Andris}},
  \bibinfo{author}{\bibfnamefont{J.}~\bibnamefont{Reades}},
  \bibinfo{author}{\bibfnamefont{M.}~\bibnamefont{Martino}},
  \bibinfo{author}{\bibfnamefont{R.}~\bibnamefont{Claxton}}, \bibnamefont{and}
  \bibinfo{author}{\bibfnamefont{S.~H.} \bibnamefont{Strogatz}},
  \bibinfo{journal}{PLoS ONE} \textbf{\bibinfo{volume}{5}},
  \bibinfo{pages}{e14248} (\bibinfo{year}{2010}),
  \urlprefix\url{http://dx.doi.org/10.1371%2Fjournal.pone.0014248}.

\bibitem[{\citenamefont{Blondel et~al.}(2010)\citenamefont{Blondel, Krings, and
  Thomas}}]{blondel2010regions}
\bibinfo{author}{\bibfnamefont{V.}~\bibnamefont{Blondel}},
  \bibinfo{author}{\bibfnamefont{G.}~\bibnamefont{Krings}}, \bibnamefont{and}
  \bibinfo{author}{\bibfnamefont{I.}~\bibnamefont{Thomas}},
  \bibinfo{journal}{Brussels Studies. La revue scientifique {\'e}lectronique
  pour les recherches sur Bruxelles/Het elektronisch wetenschappelijk
  tijdschrift voor onderzoek over Brussel/The e-journal for academic research
  on Brussels}  (\bibinfo{year}{2010}).

\bibitem[{\citenamefont{Sobolevsky et~al.}(2013)\citenamefont{Sobolevsky,
  Szell, Campari, Couronn{\'e}, Smoreda, and
  Ratti}}]{Sobolevsky2013delineating}
\bibinfo{author}{\bibfnamefont{S.}~\bibnamefont{Sobolevsky}},
  \bibinfo{author}{\bibfnamefont{M.}~\bibnamefont{Szell}},
  \bibinfo{author}{\bibfnamefont{R.}~\bibnamefont{Campari}},
  \bibinfo{author}{\bibfnamefont{T.}~\bibnamefont{Couronn{\'e}}},
  \bibinfo{author}{\bibfnamefont{Z.}~\bibnamefont{Smoreda}}, \bibnamefont{and}
  \bibinfo{author}{\bibfnamefont{C.}~\bibnamefont{Ratti}},
  \bibinfo{journal}{PloS ONE} \textbf{\bibinfo{volume}{8}},
  \bibinfo{pages}{e81707} (\bibinfo{year}{2013}).

\bibitem[{\citenamefont{Amini et~al.}(2014)\citenamefont{Amini, Kung, Kang,
  Sobolevsky, and Ratti}}]{amini2014impact}
\bibinfo{author}{\bibfnamefont{A.}~\bibnamefont{Amini}},
  \bibinfo{author}{\bibfnamefont{K.}~\bibnamefont{Kung}},
  \bibinfo{author}{\bibfnamefont{C.}~\bibnamefont{Kang}},
  \bibinfo{author}{\bibfnamefont{S.}~\bibnamefont{Sobolevsky}},
  \bibnamefont{and} \bibinfo{author}{\bibfnamefont{C.}~\bibnamefont{Ratti}},
  \bibinfo{journal}{EPJ Data Science} \textbf{\bibinfo{volume}{3}},
  \bibinfo{pages}{6} (\bibinfo{year}{2014}).

\bibitem[{\citenamefont{Hawelka et~al.}(2014)\citenamefont{Hawelka, Sitko,
  Beinat, Sobolevsky, Kazakopoulos, and Ratti}}]{hawelka2014geo}
\bibinfo{author}{\bibfnamefont{B.}~\bibnamefont{Hawelka}},
  \bibinfo{author}{\bibfnamefont{I.}~\bibnamefont{Sitko}},
  \bibinfo{author}{\bibfnamefont{E.}~\bibnamefont{Beinat}},
  \bibinfo{author}{\bibfnamefont{S.}~\bibnamefont{Sobolevsky}},
  \bibinfo{author}{\bibfnamefont{P.}~\bibnamefont{Kazakopoulos}},
  \bibnamefont{and} \bibinfo{author}{\bibfnamefont{C.}~\bibnamefont{Ratti}},
  \bibinfo{journal}{Cartography and Geographic Information Science}
  \textbf{\bibinfo{volume}{41}}, \bibinfo{pages}{260} (\bibinfo{year}{2014}).

\bibitem[{\citenamefont{Kang et~al.}(2013)\citenamefont{Kang, Sobolevsky, Liu,
  and Ratti}}]{kang2013exploring}
\bibinfo{author}{\bibfnamefont{C.}~\bibnamefont{Kang}},
  \bibinfo{author}{\bibfnamefont{S.}~\bibnamefont{Sobolevsky}},
  \bibinfo{author}{\bibfnamefont{Y.}~\bibnamefont{Liu}}, \bibnamefont{and}
  \bibinfo{author}{\bibfnamefont{C.}~\bibnamefont{Ratti}}, in
  \emph{\bibinfo{booktitle}{Proceedings of the 2nd ACM SIGKDD International
  Workshop on Urban Computing}} (\bibinfo{organization}{ACM},
  \bibinfo{year}{2013}), p.~\bibinfo{pages}{1}.

\bibitem[{\citenamefont{Sobolevsky
  et~al.}(2014{\natexlab{a}})\citenamefont{Sobolevsky, Sitko, Des~Combes,
  Hawelka, Arias, and Ratti}}]{sobolevsky2014money}
\bibinfo{author}{\bibfnamefont{S.}~\bibnamefont{Sobolevsky}},
  \bibinfo{author}{\bibfnamefont{I.}~\bibnamefont{Sitko}},
  \bibinfo{author}{\bibfnamefont{R.~T.} \bibnamefont{Des~Combes}},
  \bibinfo{author}{\bibfnamefont{B.}~\bibnamefont{Hawelka}},
  \bibinfo{author}{\bibfnamefont{J.~M.} \bibnamefont{Arias}}, \bibnamefont{and}
  \bibinfo{author}{\bibfnamefont{C.}~\bibnamefont{Ratti}}, in
  \emph{\bibinfo{booktitle}{Big Data (BigData Congress), 2014 IEEE
  International Congress on}} (\bibinfo{organization}{IEEE},
  \bibinfo{year}{2014}{\natexlab{a}}), pp. \bibinfo{pages}{136--143}.

\bibitem[{\citenamefont{Belyi et~al.}(2017)\citenamefont{Belyi, Bojic,
  Sobolevsky, Sitko, Hawelka, Rudikova, Kurbatski, and
  Ratti}}]{belyi2017global}
\bibinfo{author}{\bibfnamefont{A.}~\bibnamefont{Belyi}},
  \bibinfo{author}{\bibfnamefont{I.}~\bibnamefont{Bojic}},
  \bibinfo{author}{\bibfnamefont{S.}~\bibnamefont{Sobolevsky}},
  \bibinfo{author}{\bibfnamefont{I.}~\bibnamefont{Sitko}},
  \bibinfo{author}{\bibfnamefont{B.}~\bibnamefont{Hawelka}},
  \bibinfo{author}{\bibfnamefont{L.}~\bibnamefont{Rudikova}},
  \bibinfo{author}{\bibfnamefont{A.}~\bibnamefont{Kurbatski}},
  \bibnamefont{and} \bibinfo{author}{\bibfnamefont{C.}~\bibnamefont{Ratti}},
  \bibinfo{journal}{International Journal of Geographical Information Science}
  \textbf{\bibinfo{volume}{31}}, \bibinfo{pages}{1381} (\bibinfo{year}{2017}).

\bibitem[{\citenamefont{Grauwin et~al.}(2017)\citenamefont{Grauwin, Szell,
  Sobolevsky, H{\"o}vel, Simini, Vanhoof, Smoreda, Barab{\'a}si, and
  Ratti}}]{grauwin2017identifying}
\bibinfo{author}{\bibfnamefont{S.}~\bibnamefont{Grauwin}},
  \bibinfo{author}{\bibfnamefont{M.}~\bibnamefont{Szell}},
  \bibinfo{author}{\bibfnamefont{S.}~\bibnamefont{Sobolevsky}},
  \bibinfo{author}{\bibfnamefont{P.}~\bibnamefont{H{\"o}vel}},
  \bibinfo{author}{\bibfnamefont{F.}~\bibnamefont{Simini}},
  \bibinfo{author}{\bibfnamefont{M.}~\bibnamefont{Vanhoof}},
  \bibinfo{author}{\bibfnamefont{Z.}~\bibnamefont{Smoreda}},
  \bibinfo{author}{\bibfnamefont{A.-L.} \bibnamefont{Barab{\'a}si}},
  \bibnamefont{and} \bibinfo{author}{\bibfnamefont{C.}~\bibnamefont{Ratti}},
  \bibinfo{journal}{Scientific Reports} \textbf{\bibinfo{volume}{7}}
  (\bibinfo{year}{2017}).

\bibitem[{\citenamefont{Sobolevsky et~al.}(2018)\citenamefont{Sobolevsky, Kats,
  Malinchik, Hoffman, Kettler, and Kontokosta}}]{sobolevsky2018twitter}
\bibinfo{author}{\bibfnamefont{S.}~\bibnamefont{Sobolevsky}},
  \bibinfo{author}{\bibfnamefont{P.}~\bibnamefont{Kats}},
  \bibinfo{author}{\bibfnamefont{S.}~\bibnamefont{Malinchik}},
  \bibinfo{author}{\bibfnamefont{M.}~\bibnamefont{Hoffman}},
  \bibinfo{author}{\bibfnamefont{B.}~\bibnamefont{Kettler}}, \bibnamefont{and}
  \bibinfo{author}{\bibfnamefont{C.}~\bibnamefont{Kontokosta}}, in
  \emph{\bibinfo{booktitle}{Proceedings of the 51st Hawaii International
  Conference on System Sciences}} (\bibinfo{year}{2018}).

\bibitem[{\citenamefont{Landsman et~al.}(2020)\citenamefont{Landsman, Kats,
  Nenko, and Sobolevsky}}]{landsman2020zoning}
\bibinfo{author}{\bibfnamefont{D.}~\bibnamefont{Landsman}},
  \bibinfo{author}{\bibfnamefont{P.}~\bibnamefont{Kats}},
  \bibinfo{author}{\bibfnamefont{A.}~\bibnamefont{Nenko}}, \bibnamefont{and}
  \bibinfo{author}{\bibfnamefont{S.}~\bibnamefont{Sobolevsky}},
  \bibinfo{journal}{Procedia Computer Science} \textbf{\bibinfo{volume}{178}},
  \bibinfo{pages}{125} (\bibinfo{year}{2020}).

\bibitem[{\citenamefont{Landsman et~al.}(2021)\citenamefont{Landsman, Kats,
  Nenko, Kudinov, and Sobolevsky}}]{landsman2021social}
\bibinfo{author}{\bibfnamefont{D.}~\bibnamefont{Landsman}},
  \bibinfo{author}{\bibfnamefont{P.}~\bibnamefont{Kats}},
  \bibinfo{author}{\bibfnamefont{A.}~\bibnamefont{Nenko}},
  \bibinfo{author}{\bibfnamefont{S.}~\bibnamefont{Kudinov}}, \bibnamefont{and}
  \bibinfo{author}{\bibfnamefont{S.}~\bibnamefont{Sobolevsky}}, in
  \emph{\bibinfo{booktitle}{Proceedings of the 54th Hawaii International
  Conference on System Sciences}} (\bibinfo{year}{2021}), p.
  \bibinfo{pages}{1149}.

\bibitem[{\citenamefont{Hastie}(2001)}]{Hastie2001ElementsOfStatisticalLearning}
\bibinfo{author}{\bibfnamefont{T.}~\bibnamefont{Hastie}},
  \emph{\bibinfo{title}{The elements of statistical learning : data mining,
  inference, and prediction : with 200 full-color illustrations}}
  (\bibinfo{publisher}{Springer}, \bibinfo{address}{New York},
  \bibinfo{year}{2001}), ISBN \bibinfo{isbn}{0387952845}.

\bibitem[{\citenamefont{Girvan and Newman}(2002{\natexlab{a}})}]{GN}
\bibinfo{author}{\bibfnamefont{M.}~\bibnamefont{Girvan}} \bibnamefont{and}
  \bibinfo{author}{\bibfnamefont{M.}~\bibnamefont{Newman}},
  \bibinfo{journal}{Proc. Natl. Acad. Sci. USA} \textbf{\bibinfo{volume}{99
  (12)}}, \bibinfo{pages}{7821} (\bibinfo{year}{2002}{\natexlab{a}}).

\bibitem[{\citenamefont{Newman and Girvan}(2004)}]{newman2004}
\bibinfo{author}{\bibfnamefont{M.}~\bibnamefont{Newman}} \bibnamefont{and}
  \bibinfo{author}{\bibfnamefont{M.}~\bibnamefont{Girvan}},
  \bibinfo{journal}{Phys. Rev. E} \textbf{\bibinfo{volume}{69 (2)}},
  \bibinfo{pages}{026113} (\bibinfo{year}{2004}).

\bibitem[{\citenamefont{Newman}(2006{\natexlab{a}})}]{newman2006}
\bibinfo{author}{\bibfnamefont{M.}~\bibnamefont{Newman}},
  \bibinfo{journal}{Proceedings of the National Academy of Sciences}
  \textbf{\bibinfo{volume}{103}}, \bibinfo{pages}{8577}
  (\bibinfo{year}{2006}{\natexlab{a}}).

\bibitem[{\citenamefont{Newman}(2004)}]{NewmanPRE2004}
\bibinfo{author}{\bibfnamefont{M.~E.~J.} \bibnamefont{Newman}},
  \bibinfo{journal}{Phys. Rev. E} \textbf{\bibinfo{volume}{69}},
  \bibinfo{pages}{066133} (\bibinfo{year}{2004}),
  \urlprefix\url{http://link.aps.org/doi/10.1103/PhysRevE.69.066133}.

\bibitem[{\citenamefont{Clauset et~al.}(2004)\citenamefont{Clauset, Newman, and
  Moore}}]{CNM2004VeryLargeNetworks}
\bibinfo{author}{\bibfnamefont{A.}~\bibnamefont{Clauset}},
  \bibinfo{author}{\bibfnamefont{M.~E.~J.} \bibnamefont{Newman}},
  \bibnamefont{and} \bibinfo{author}{\bibfnamefont{C.}~\bibnamefont{Moore}},
  \bibinfo{journal}{Phys. Rev. E} \textbf{\bibinfo{volume}{70}},
  \bibinfo{pages}{066111} (\bibinfo{year}{2004}),
  \urlprefix\url{http://link.aps.org/doi/10.1103/PhysRevE.70.066111}.

\bibitem[{\citenamefont{Sun et~al.}(2009)\citenamefont{Sun, Danila, Josi{\'c},
  and Bassler}}]{Sun2009}
\bibinfo{author}{\bibfnamefont{Y.}~\bibnamefont{Sun}},
  \bibinfo{author}{\bibfnamefont{B.}~\bibnamefont{Danila}},
  \bibinfo{author}{\bibfnamefont{K.}~\bibnamefont{Josi{\'c}}},
  \bibnamefont{and} \bibinfo{author}{\bibfnamefont{K.~E.}
  \bibnamefont{Bassler}}, \bibinfo{journal}{EPL (Europhysics Letters)}
  \textbf{\bibinfo{volume}{86}}, \bibinfo{pages}{28004} (\bibinfo{year}{2009}),
  \urlprefix\url{http://stacks.iop.org/0295-5075/86/i=2/a=28004}.

\bibitem[{\citenamefont{Blondel et~al.}(2008)\citenamefont{Blondel, Guillaume,
  Lambiotte, and Lefebvre}}]{louvain}
\bibinfo{author}{\bibfnamefont{V.~D.} \bibnamefont{Blondel}},
  \bibinfo{author}{\bibfnamefont{J.-L.} \bibnamefont{Guillaume}},
  \bibinfo{author}{\bibfnamefont{R.}~\bibnamefont{Lambiotte}},
  \bibnamefont{and} \bibinfo{author}{\bibfnamefont{E.}~\bibnamefont{Lefebvre}},
  \bibinfo{journal}{Journal of Statistical Mechanics: Theory and Experiment}
  \textbf{\bibinfo{volume}{2008}}, \bibinfo{pages}{P10008}
  (\bibinfo{year}{2008}).

\bibitem[{\citenamefont{Guimera et~al.}(2004)\citenamefont{Guimera,
  Sales-Pardo, and Amaral}}]{simulatedAnnealing}
\bibinfo{author}{\bibfnamefont{R.}~\bibnamefont{Guimera}},
  \bibinfo{author}{\bibfnamefont{M.}~\bibnamefont{Sales-Pardo}},
  \bibnamefont{and} \bibinfo{author}{\bibfnamefont{L.~A.~N.}
  \bibnamefont{Amaral}}, \bibinfo{journal}{Physical Review E}
  \textbf{\bibinfo{volume}{70}}, \bibinfo{pages}{025101}
  (\bibinfo{year}{2004}).

\bibitem[{\citenamefont{Good et~al.}(2010)\citenamefont{Good, de~Montjoye, and
  Clauset}}]{Good2010PerformanceOfModularity}
\bibinfo{author}{\bibfnamefont{B.~H.} \bibnamefont{Good}},
  \bibinfo{author}{\bibfnamefont{Y.-A.} \bibnamefont{de~Montjoye}},
  \bibnamefont{and} \bibinfo{author}{\bibfnamefont{A.}~\bibnamefont{Clauset}},
  \bibinfo{journal}{Phys. Rev. E} \textbf{\bibinfo{volume}{81}},
  \bibinfo{pages}{046106} (\bibinfo{year}{2010}),
  \urlprefix\url{http://link.aps.org/doi/10.1103/PhysRevE.81.046106}.

\bibitem[{\citenamefont{Duch and Arenas}(2005)}]{Duch2005CElegans}
\bibinfo{author}{\bibfnamefont{J.}~\bibnamefont{Duch}} \bibnamefont{and}
  \bibinfo{author}{\bibfnamefont{A.}~\bibnamefont{Arenas}},
  \bibinfo{journal}{Phys. Rev. E} \textbf{\bibinfo{volume}{72}},
  \bibinfo{pages}{027104} (\bibinfo{year}{2005}),
  \urlprefix\url{http://link.aps.org/doi/10.1103/PhysRevE.72.027104}.

\bibitem[{\citenamefont{Lee et~al.}(2012)\citenamefont{Lee, Gross, and
  Lee}}]{LeeCSA}
\bibinfo{author}{\bibfnamefont{J.}~\bibnamefont{Lee}},
  \bibinfo{author}{\bibfnamefont{S.~P.} \bibnamefont{Gross}}, \bibnamefont{and}
  \bibinfo{author}{\bibfnamefont{J.}~\bibnamefont{Lee}},
  \bibinfo{journal}{Phys. Rev. E} \textbf{\bibinfo{volume}{85}},
  \bibinfo{pages}{056702} (\bibinfo{year}{2012}),
  \urlprefix\url{http://link.aps.org/doi/10.1103/PhysRevE.85.056702}.

\bibitem[{\citenamefont{Sobolevsky
  et~al.}(2014{\natexlab{b}})\citenamefont{Sobolevsky, Campari, Belyi, and
  Ratti}}]{combo}
\bibinfo{author}{\bibfnamefont{S.}~\bibnamefont{Sobolevsky}},
  \bibinfo{author}{\bibfnamefont{R.}~\bibnamefont{Campari}},
  \bibinfo{author}{\bibfnamefont{A.}~\bibnamefont{Belyi}}, \bibnamefont{and}
  \bibinfo{author}{\bibfnamefont{C.}~\bibnamefont{Ratti}},
  \bibinfo{journal}{Physical Review E} \textbf{\bibinfo{volume}{90}},
  \bibinfo{pages}{012811} (\bibinfo{year}{2014}{\natexlab{b}}).

\bibitem[{\citenamefont{Fortunato}(2010)}]{fortunato2010}
\bibinfo{author}{\bibfnamefont{S.}~\bibnamefont{Fortunato}},
  \bibinfo{journal}{Physics Report} \textbf{\bibinfo{volume}{486}},
  \bibinfo{pages}{75} (\bibinfo{year}{2010}).

\bibitem[{\citenamefont{Fortunato and Hric}(2016)}]{fortunato2016community}
\bibinfo{author}{\bibfnamefont{S.}~\bibnamefont{Fortunato}} \bibnamefont{and}
  \bibinfo{author}{\bibfnamefont{D.}~\bibnamefont{Hric}},
  \bibinfo{journal}{Physics reports} \textbf{\bibinfo{volume}{659}},
  \bibinfo{pages}{1} (\bibinfo{year}{2016}).

\bibitem[{\citenamefont{Khan and Niazi}(2017)}]{khan2017network}
\bibinfo{author}{\bibfnamefont{B.~S.} \bibnamefont{Khan}} \bibnamefont{and}
  \bibinfo{author}{\bibfnamefont{M.~A.} \bibnamefont{Niazi}},
  \bibinfo{journal}{arXiv preprint arXiv:1708.00977}  (\bibinfo{year}{2017}).

\bibitem[{\citenamefont{Javed et~al.}(2018)\citenamefont{Javed, Younis, Latif,
  Qadir, and Baig}}]{javed2018community}
\bibinfo{author}{\bibfnamefont{M.~A.} \bibnamefont{Javed}},
  \bibinfo{author}{\bibfnamefont{M.~S.} \bibnamefont{Younis}},
  \bibinfo{author}{\bibfnamefont{S.}~\bibnamefont{Latif}},
  \bibinfo{author}{\bibfnamefont{J.}~\bibnamefont{Qadir}}, \bibnamefont{and}
  \bibinfo{author}{\bibfnamefont{A.}~\bibnamefont{Baig}},
  \bibinfo{journal}{Journal of Network and Computer Applications}
  \textbf{\bibinfo{volume}{108}}, \bibinfo{pages}{87} (\bibinfo{year}{2018}).

\bibitem[{\citenamefont{Brandes et~al.}(2006)\citenamefont{Brandes, Delling,
  Gaertler, G{\"o}rke, Hoefer, Nikoloski, and Wagner}}]{brandes2006maximizing}
\bibinfo{author}{\bibfnamefont{U.}~\bibnamefont{Brandes}},
  \bibinfo{author}{\bibfnamefont{D.}~\bibnamefont{Delling}},
  \bibinfo{author}{\bibfnamefont{M.}~\bibnamefont{Gaertler}},
  \bibinfo{author}{\bibfnamefont{R.}~\bibnamefont{G{\"o}rke}},
  \bibinfo{author}{\bibfnamefont{M.}~\bibnamefont{Hoefer}},
  \bibinfo{author}{\bibfnamefont{Z.}~\bibnamefont{Nikoloski}},
  \bibnamefont{and} \bibinfo{author}{\bibfnamefont{D.}~\bibnamefont{Wagner}},
  \bibinfo{journal}{arXiv preprint physics/0608255}  (\bibinfo{year}{2006}).

\bibitem[{\citenamefont{Sobolevsky et~al.}(2017)\citenamefont{Sobolevsky,
  Belyi, and Ratti}}]{sobolevsky2017optimality}
\bibinfo{author}{\bibfnamefont{S.}~\bibnamefont{Sobolevsky}},
  \bibinfo{author}{\bibfnamefont{A.}~\bibnamefont{Belyi}}, \bibnamefont{and}
  \bibinfo{author}{\bibfnamefont{C.}~\bibnamefont{Ratti}},
  \bibinfo{journal}{arXiv preprint arXiv:1712.05110}  (\bibinfo{year}{2017}).

\bibitem[{\citenamefont{Wu et~al.}(2020)\citenamefont{Wu, Pan, Chen, Long,
  Zhang, and Philip}}]{wu2020comprehensive}
\bibinfo{author}{\bibfnamefont{Z.}~\bibnamefont{Wu}},
  \bibinfo{author}{\bibfnamefont{S.}~\bibnamefont{Pan}},
  \bibinfo{author}{\bibfnamefont{F.}~\bibnamefont{Chen}},
  \bibinfo{author}{\bibfnamefont{G.}~\bibnamefont{Long}},
  \bibinfo{author}{\bibfnamefont{C.}~\bibnamefont{Zhang}}, \bibnamefont{and}
  \bibinfo{author}{\bibfnamefont{S.~Y.} \bibnamefont{Philip}},
  \bibinfo{journal}{IEEE transactions on neural networks and learning systems}
  (\bibinfo{year}{2020}).

\bibitem[{\citenamefont{Chen et~al.}(2017)\citenamefont{Chen, Li, and
  Bruna}}]{chen2017supervised}
\bibinfo{author}{\bibfnamefont{Z.}~\bibnamefont{Chen}},
  \bibinfo{author}{\bibfnamefont{X.}~\bibnamefont{Li}}, \bibnamefont{and}
  \bibinfo{author}{\bibfnamefont{J.}~\bibnamefont{Bruna}},
  \bibinfo{journal}{arXiv preprint arXiv:1705.08415}  (\bibinfo{year}{2017}).

\bibitem[{\citenamefont{Bruna and Li}(2017)}]{bruna2017community}
\bibinfo{author}{\bibfnamefont{J.}~\bibnamefont{Bruna}} \bibnamefont{and}
  \bibinfo{author}{\bibfnamefont{X.}~\bibnamefont{Li}}, \bibinfo{journal}{stat}
  \textbf{\bibinfo{volume}{1050}}, \bibinfo{pages}{27} (\bibinfo{year}{2017}).

\bibitem[{\citenamefont{Shchur and
  G{\"u}nnemann}(2019)}]{shchur2019overlapping}
\bibinfo{author}{\bibfnamefont{O.}~\bibnamefont{Shchur}} \bibnamefont{and}
  \bibinfo{author}{\bibfnamefont{S.}~\bibnamefont{G{\"u}nnemann}},
  \bibinfo{journal}{arXiv preprint arXiv:1909.12201}  (\bibinfo{year}{2019}).

\bibitem[{\citenamefont{Bandyopadhyay and Peter}(2020)}]{bandyopadhyay2020self}
\bibinfo{author}{\bibfnamefont{S.}~\bibnamefont{Bandyopadhyay}}
  \bibnamefont{and} \bibinfo{author}{\bibfnamefont{V.}~\bibnamefont{Peter}},
  \bibinfo{journal}{arXiv preprint arXiv:2011.14078}  (\bibinfo{year}{2020}).

\bibitem[{\citenamefont{Fortunato and
  Barth{\'e}l{\'e}my}(2007)}]{Fortunato02012007ResolutionLimit}
\bibinfo{author}{\bibfnamefont{S.}~\bibnamefont{Fortunato}} \bibnamefont{and}
  \bibinfo{author}{\bibfnamefont{M.}~\bibnamefont{Barth{\'e}l{\'e}my}},
  \bibinfo{journal}{Proceedings of the National Academy of Sciences}
  \textbf{\bibinfo{volume}{104}}, \bibinfo{pages}{36} (\bibinfo{year}{2007}),
  \eprint{http://www.pnas.org/content/104/1/36.full.pdf+html},
  \urlprefix\url{http://www.pnas.org/content/104/1/36.abstract}.

\bibitem[{\citenamefont{Rosvall and
  Bergstrom}(2007)}]{Rosvall01052007InformationTheoretic}
\bibinfo{author}{\bibfnamefont{M.}~\bibnamefont{Rosvall}} \bibnamefont{and}
  \bibinfo{author}{\bibfnamefont{C.~T.} \bibnamefont{Bergstrom}},
  \bibinfo{journal}{Proceedings of the National Academy of Sciences}
  \textbf{\bibinfo{volume}{104}}, \bibinfo{pages}{7327} (\bibinfo{year}{2007}),
  \eprint{http://www.pnas.org/content/104/18/7327.full.pdf+html},
  \urlprefix\url{http://www.pnas.org/content/104/18/7327.abstract}.

\bibitem[{\citenamefont{Rosvall and Bergstrom}(2008)}]{Infomap}
\bibinfo{author}{\bibfnamefont{M.}~\bibnamefont{Rosvall}} \bibnamefont{and}
  \bibinfo{author}{\bibfnamefont{C.}~\bibnamefont{Bergstrom}},
  \bibinfo{journal}{Proc. Natl. Acad. Sci. USA} \textbf{\bibinfo{volume}{105}},
  \bibinfo{pages}{1118} (\bibinfo{year}{2008}).

\bibitem[{\citenamefont{Karrer and Newman}(2011)}]{Newman2011Stochastic}
\bibinfo{author}{\bibfnamefont{B.}~\bibnamefont{Karrer}} \bibnamefont{and}
  \bibinfo{author}{\bibfnamefont{M.~E.~J.} \bibnamefont{Newman}},
  \bibinfo{journal}{Phys. Rev. E} \textbf{\bibinfo{volume}{83}},
  \bibinfo{pages}{016107} (\bibinfo{year}{2011}),
  \urlprefix\url{http://link.aps.org/doi/10.1103/PhysRevE.83.016107}.

\bibitem[{\citenamefont{Ball et~al.}(2011)\citenamefont{Ball, Karrer, and
  Newman}}]{Newman2011Efficient}
\bibinfo{author}{\bibfnamefont{B.}~\bibnamefont{Ball}},
  \bibinfo{author}{\bibfnamefont{B.}~\bibnamefont{Karrer}}, \bibnamefont{and}
  \bibinfo{author}{\bibfnamefont{M.~E.~J.} \bibnamefont{Newman}},
  \bibinfo{journal}{Phys. Rev. E} \textbf{\bibinfo{volume}{84}},
  \bibinfo{pages}{036103} (\bibinfo{year}{2011}),
  \urlprefix\url{http://link.aps.org/doi/10.1103/PhysRevE.84.036103}.

\bibitem[{\citenamefont{Bickel and Chen}(2009)}]{Bickel2009Nonparametric}
\bibinfo{author}{\bibfnamefont{P.~J.} \bibnamefont{Bickel}} \bibnamefont{and}
  \bibinfo{author}{\bibfnamefont{A.}~\bibnamefont{Chen}},
  \bibinfo{journal}{Proceedings of the National Academy of Sciences}
  \textbf{\bibinfo{volume}{106}}, \bibinfo{pages}{21068}
  (\bibinfo{year}{2009}).

\bibitem[{\citenamefont{Decelle
  et~al.}(2011{\natexlab{a}})\citenamefont{Decelle, Krzakala, Moore, and
  Zdeborov\'a}}]{Decelle2011BlockModel}
\bibinfo{author}{\bibfnamefont{A.}~\bibnamefont{Decelle}},
  \bibinfo{author}{\bibfnamefont{F.}~\bibnamefont{Krzakala}},
  \bibinfo{author}{\bibfnamefont{C.}~\bibnamefont{Moore}}, \bibnamefont{and}
  \bibinfo{author}{\bibfnamefont{L.}~\bibnamefont{Zdeborov\'a}},
  \bibinfo{journal}{Phys. Rev. Lett.} \textbf{\bibinfo{volume}{107}},
  \bibinfo{pages}{065701} (\bibinfo{year}{2011}{\natexlab{a}}),
  \urlprefix\url{http://link.aps.org/doi/10.1103/PhysRevLett.107.065701}.

\bibitem[{\citenamefont{Decelle
  et~al.}(2011{\natexlab{b}})\citenamefont{Decelle, Krzakala, Moore, and
  Zdeborov\'a}}]{Decelle2011BlockModelAsymptotics}
\bibinfo{author}{\bibfnamefont{A.}~\bibnamefont{Decelle}},
  \bibinfo{author}{\bibfnamefont{F.}~\bibnamefont{Krzakala}},
  \bibinfo{author}{\bibfnamefont{C.}~\bibnamefont{Moore}}, \bibnamefont{and}
  \bibinfo{author}{\bibfnamefont{L.}~\bibnamefont{Zdeborov\'a}},
  \bibinfo{journal}{Phys. Rev. E} \textbf{\bibinfo{volume}{84}},
  \bibinfo{pages}{066106} (\bibinfo{year}{2011}{\natexlab{b}}),
  \urlprefix\url{http://link.aps.org/doi/10.1103/PhysRevE.84.066106}.

\bibitem[{\citenamefont{Yan et~al.}(2014)\citenamefont{Yan, Shalizi, Jensen,
  Krzakala, Moore, Zdeborov{\'a}, Zhang, and Zhu}}]{Yan2012ModelSelection}
\bibinfo{author}{\bibfnamefont{X.}~\bibnamefont{Yan}},
  \bibinfo{author}{\bibfnamefont{C.}~\bibnamefont{Shalizi}},
  \bibinfo{author}{\bibfnamefont{J.~E.} \bibnamefont{Jensen}},
  \bibinfo{author}{\bibfnamefont{F.}~\bibnamefont{Krzakala}},
  \bibinfo{author}{\bibfnamefont{C.}~\bibnamefont{Moore}},
  \bibinfo{author}{\bibfnamefont{L.}~\bibnamefont{Zdeborov{\'a}}},
  \bibinfo{author}{\bibfnamefont{P.}~\bibnamefont{Zhang}}, \bibnamefont{and}
  \bibinfo{author}{\bibfnamefont{Y.}~\bibnamefont{Zhu}},
  \bibinfo{journal}{Journal of Statistical Mechanics: Theory and Experiment}
  \textbf{\bibinfo{volume}{2014}}, \bibinfo{pages}{P05007}
  (\bibinfo{year}{2014}).

\bibitem[{\citenamefont{Aldecoa and Mar\`in}(2011)}]{Aldecoa2011Deciphering}
\bibinfo{author}{\bibfnamefont{R.}~\bibnamefont{Aldecoa}} \bibnamefont{and}
  \bibinfo{author}{\bibfnamefont{I.}~\bibnamefont{Mar\`in}},
  \bibinfo{journal}{PLoS ONE} \textbf{\bibinfo{volume}{6}},
  \bibinfo{pages}{e24195} (\bibinfo{year}{2011}),
  \urlprefix\url{http://dx.doi.org/10.1371%2Fjournal.pone.0024195}.

\bibitem[{\citenamefont{Weisfeiler and Leman}(1968)}]{weisfeiler1968reduction}
\bibinfo{author}{\bibfnamefont{B.}~\bibnamefont{Weisfeiler}} \bibnamefont{and}
  \bibinfo{author}{\bibfnamefont{A.}~\bibnamefont{Leman}},
  \bibinfo{journal}{NTI, Series} \textbf{\bibinfo{volume}{2}},
  \bibinfo{pages}{12} (\bibinfo{year}{1968}).

\bibitem[{\citenamefont{Lancichinetti et~al.}(2008)\citenamefont{Lancichinetti,
  Fortunato, and Radicchi}}]{LFR}
\bibinfo{author}{\bibfnamefont{A.}~\bibnamefont{Lancichinetti}},
  \bibinfo{author}{\bibfnamefont{S.}~\bibnamefont{Fortunato}},
  \bibnamefont{and} \bibinfo{author}{\bibfnamefont{F.}~\bibnamefont{Radicchi}},
  \bibinfo{journal}{Phys. Rev. E} \textbf{\bibinfo{volume}{78 (4)}},
  \bibinfo{pages}{046110} (\bibinfo{year}{2008}).

\bibitem[{\citenamefont{Holland et~al.}(1983)\citenamefont{Holland, Laskey, and
  Leinhardt}}]{holland1983stochastic}
\bibinfo{author}{\bibfnamefont{P.~W.} \bibnamefont{Holland}},
  \bibinfo{author}{\bibfnamefont{K.~B.} \bibnamefont{Laskey}},
  \bibnamefont{and}
  \bibinfo{author}{\bibfnamefont{S.}~\bibnamefont{Leinhardt}},
  \bibinfo{journal}{Social networks} \textbf{\bibinfo{volume}{5}},
  \bibinfo{pages}{109} (\bibinfo{year}{1983}).

\bibitem[{\citenamefont{Ma et~al.}(2020)\citenamefont{Ma, Guo, Ren, Tang, and
  Yin}}]{ma2020streaming}
\bibinfo{author}{\bibfnamefont{Y.}~\bibnamefont{Ma}},
  \bibinfo{author}{\bibfnamefont{Z.}~\bibnamefont{Guo}},
  \bibinfo{author}{\bibfnamefont{Z.}~\bibnamefont{Ren}},
  \bibinfo{author}{\bibfnamefont{J.}~\bibnamefont{Tang}}, \bibnamefont{and}
  \bibinfo{author}{\bibfnamefont{D.}~\bibnamefont{Yin}}, in
  \emph{\bibinfo{booktitle}{Proceedings of the 43rd International ACM SIGIR
  Conference on Research and Development in Information Retrieval}}
  (\bibinfo{year}{2020}), pp. \bibinfo{pages}{719--728}.

\bibitem[{\citenamefont{Zachary}(1977)}]{zachary1977ifm}
\bibinfo{author}{\bibfnamefont{W.~W.} \bibnamefont{Zachary}},
  \bibinfo{journal}{Journal of Anthropological Research}
  \textbf{\bibinfo{volume}{33}}, \bibinfo{pages}{452} (\bibinfo{year}{1977}).

\bibitem[{\citenamefont{Lusseau et~al.}(2003)\citenamefont{Lusseau, Schneider,
  Boisseau, Haase, Slooten, and Dawson}}]{Lusseau2003Dolphins}
\bibinfo{author}{\bibfnamefont{D.}~\bibnamefont{Lusseau}},
  \bibinfo{author}{\bibfnamefont{K.}~\bibnamefont{Schneider}},
  \bibinfo{author}{\bibfnamefont{O.~J.} \bibnamefont{Boisseau}},
  \bibinfo{author}{\bibfnamefont{P.}~\bibnamefont{Haase}},
  \bibinfo{author}{\bibfnamefont{E.}~\bibnamefont{Slooten}}, \bibnamefont{and}
  \bibinfo{author}{\bibfnamefont{S.~M.} \bibnamefont{Dawson}},
  \bibinfo{journal}{Behavioral Ecology and Sociobiology}
  \textbf{\bibinfo{volume}{54}}, \bibinfo{pages}{396} (\bibinfo{year}{2003}),
  \urlprefix\url{http://dx.doi.org/10.1007/s00265-003-0651-y}.

\bibitem[{\citenamefont{Knuth}(1993)}]{Knuth1993GraphBase}
\bibinfo{author}{\bibfnamefont{D.~E.} \bibnamefont{Knuth}},
  \emph{\bibinfo{title}{{The Stanford GraphBase: a platform for combinatorial
  computing}}} (\bibinfo{publisher}{Addison-Wesley}, \bibinfo{year}{1993}),
  \urlprefix\url{http://www-cs-staff.stanford.edu/\~{}uno/sgb.html}.

\bibitem[{\citenamefont{Girvan and
  Newman}(2002{\natexlab{b}})}]{Girvan11062002CommStruct}
\bibinfo{author}{\bibfnamefont{M.}~\bibnamefont{Girvan}} \bibnamefont{and}
  \bibinfo{author}{\bibfnamefont{M.~E.~J.} \bibnamefont{Newman}},
  \bibinfo{journal}{Proceedings of the National Academy of Sciences}
  \textbf{\bibinfo{volume}{99}}, \bibinfo{pages}{7821}
  (\bibinfo{year}{2002}{\natexlab{b}}),
  \eprint{http://www.pnas.org/content/99/12/7821.full.pdf+html},
  \urlprefix\url{http://www.pnas.org/content/99/12/7821.abstract}.

\bibitem[{\citenamefont{Newman}(2006{\natexlab{b}})}]{Newman2006CommunityStructureEigen}
\bibinfo{author}{\bibfnamefont{M.~E.~J.} \bibnamefont{Newman}},
  \bibinfo{journal}{Phys. Rev. E} \textbf{\bibinfo{volume}{74}},
  \bibinfo{pages}{036104} (\bibinfo{year}{2006}{\natexlab{b}}),
  \urlprefix\url{http://link.aps.org/doi/10.1103/PhysRevE.74.036104}.

\bibitem[{\citenamefont{Gleiser and Danon}(2003)}]{Gleiser2003Jazz}
\bibinfo{author}{\bibfnamefont{P.~M.} \bibnamefont{Gleiser}} \bibnamefont{and}
  \bibinfo{author}{\bibfnamefont{L.}~\bibnamefont{Danon}},
  \bibinfo{journal}{Advances in Complex Systems} \textbf{\bibinfo{volume}{06}},
  \bibinfo{pages}{565} (\bibinfo{year}{2003}),
  \eprint{http://www.worldscientific.com/doi/pdf/10.1142/S0219525903001067},
  \urlprefix\url{http://www.worldscientific.com/doi/abs/10.1142/S0219525903001067}.

\bibitem[{\citenamefont{White et~al.}(1986)\citenamefont{White, Southgate,
  Thomson, and Brenner}}]{White12111986}
\bibinfo{author}{\bibfnamefont{J.~G.} \bibnamefont{White}},
  \bibinfo{author}{\bibfnamefont{E.}~\bibnamefont{Southgate}},
  \bibinfo{author}{\bibfnamefont{J.~N.} \bibnamefont{Thomson}},
  \bibnamefont{and} \bibinfo{author}{\bibfnamefont{S.}~\bibnamefont{Brenner}},
  \bibinfo{journal}{Philosophical Transactions of the Royal Society of London.
  B, Biological Sciences} \textbf{\bibinfo{volume}{314}}, \bibinfo{pages}{1}
  (\bibinfo{year}{1986}),
  \eprint{http://rstb.royalsocietypublishing.org/content/314/1165/1.full.pdf+html},
  \urlprefix\url{http://rstb.royalsocietypublishing.org/content/314/1165/1.abstract}.

\bibitem[{\citenamefont{Guimer\`a et~al.}(2003)\citenamefont{Guimer\`a, Danon,
  D\'iaz-Guilera, Giralt, and Arenas}}]{Guimera2003Email}
\bibinfo{author}{\bibfnamefont{R.}~\bibnamefont{Guimer\`a}},
  \bibinfo{author}{\bibfnamefont{L.}~\bibnamefont{Danon}},
  \bibinfo{author}{\bibfnamefont{A.}~\bibnamefont{D\'iaz-Guilera}},
  \bibinfo{author}{\bibfnamefont{F.}~\bibnamefont{Giralt}}, \bibnamefont{and}
  \bibinfo{author}{\bibfnamefont{A.}~\bibnamefont{Arenas}},
  \bibinfo{journal}{Phys. Rev. E} \textbf{\bibinfo{volume}{68}},
  \bibinfo{pages}{065103} (\bibinfo{year}{2003}),
  \urlprefix\url{http://link.aps.org/doi/10.1103/PhysRevE.68.065103}.

\bibitem[{\citenamefont{Adamic and
  Glance}(2005)}]{Adamic2005PoliticalBlogsphere}
\bibinfo{author}{\bibfnamefont{L.~A.} \bibnamefont{Adamic}} \bibnamefont{and}
  \bibinfo{author}{\bibfnamefont{N.}~\bibnamefont{Glance}}, in
  \emph{\bibinfo{booktitle}{Proceedings of the 3rd international workshop on
  Link discovery}} (\bibinfo{publisher}{ACM}, \bibinfo{address}{New York, NY,
  USA}, \bibinfo{year}{2005}), LinkKDD '05, pp. \bibinfo{pages}{36--43}, ISBN
  \bibinfo{isbn}{1-59593-215-1},
  \urlprefix\url{http://doi.acm.org/10.1145/1134271.1134277}.

\end{thebibliography}

\section{Appendix. Sources and characteristics of the classic network examples}

The analysis presented in the paper includes 12 classic sample networks. Their sources and characteristics (network size and whether the network is weighted and/or directed) are presented in the table \ref{tab:networksList}.

%match the table with results (Neural/Metabolic!)

\begin{table*}[htpb]
\caption{\label{tab:networksList}List, with sources, of the networks we used in our benchmark.}

\begin{center}
\begin{tabular}{| c | p{2.5cm} | p{7cm} | c | c | c |}
\hline
\textbf{No} & \textbf{Name} & \textbf{Description} & \textbf{Nodes} & \textbf{Weighted} & \textbf{Directed} \\
\hline
1 & Karate & Zachary's Karate network \cite{zachary1977ifm} & 34 & NO & NO\\
2 & Dolphins & Dolphins' Social Network \cite{Lusseau2003Dolphins} & 62 & NO & NO\\
3 & Les Miserable & Coappeareance of characters in Les Miserable \cite{Knuth1993GraphBase} & 77 & NO & YES\\
4 & Football & American College Football games in year 2000\cite{Girvan11062002CommStruct} & 115 & NO & YES\\
5 & Political Books & Amazon.com Co-purchases of political books\textsuperscript{2} & 105 & NO & NO\\
6 & Copperfield & Common adjective and noun adjacencies in David Copperfield\cite{Newman2006CommunityStructureEigen} & 112 & NO & NO \\

7 & Jazz & Network of Jazz Musicians\cite{Gleiser2003Jazz} & 198 & YES & YES\\

8 & C. Elegans & Neural network of C. Elegans\cite{White12111986} & 297 & YES & YES\\

9 & Airports97 & US Aiports network from 1997\textsuperscript{3} & 332 & YES & YES \\

%9 & Metabolic & Metabolic Network of C. Elegans\cite{Duch2005CElegans} & 453 & YES & YES\\

10 & Email & Email Networks University of Tarragona\cite{Guimera2003Email} & 1133 & NO & NO\\

11 & Blogs & Connections among political blogs\cite{Adamic2005PoliticalBlogsphere} & 1490 & YES & YES\\
%9 & Coauthorship & Coauthorship in network science\cite{Newman2006CommunityStructureEigen} & 1589 & YES & YES\\
12 & Airports2010 & Complete network of US airports in 2010\textsuperscript{4} & 1858 & YES & YES\\
%18 & 410 & Network extracted from the Infectious: STAY AWAY exhibition\cite{Isella2011166Infectious}\\

%10 & 2114 & Protein interaction network for Saccharomyces Cerevisiae\cite{Jeong2001}\\
%16 & 8297 & Wiki vote network\cite{Leskovec2010SignedNetworks}\\

\hline
\end{tabular}
\\[10pt]
\end{center}
\end{table*}

\footnotetext{Valdis Krebs, data available online at http://www.orgnet.com}
\footnotetext{Vladimir Batagelj and Andrej Mrvar (2006): Pajek datasets. Airports. http://vlado.fmf.uni-lj.si/pub/networks/data/mix/USAir97.net}
\footnotetext{data from the Bureau of Transportation Statistics - details at http://toreopsahl.com/datasets/\#usairports}

Five of the 12 sample networks are directed in their original form. As available to us Python NetworkX implementation of the Louvain method only handles undirected network we have summarized the networks for the purpose of the comparative evaluation of all the three partition methods. However Combo and GNNS implementations are capable of handling directed versions of the networks and in the table \ref{tab:directedMod} below we present the achieved modularity scores by Combo, GNNS100 and GNNS2500 for the original directed versions of the sample networks. One can see that except of Jazz, where Combo underperforms, all three methods reach closely similar modularity scores with a slight lead of Combo.

\begin{table*}[htpb]
\caption{\label{tab:directedMod}Achieved modularity scores for the original directed configurations of the classical network examples.}

\begin{center}
\begin{tabular}{| c | p{3cm} | p{2cm} | p{2cm} | p{2cm} |}
\hline
\textbf{No} & \textbf{Name} & \multicolumn{3}{c|}{best modularity score} \\
\cline{3-5}
& & Combo & GNNS 100 & GNNS 1000\\
\hline
1 & Jazz & 0.444787 & 0.445550 & 0.445627 \\
2 & C. Elegans & 0.507642 & 0.507642 & 0.507642 \\
3 & Airports97 & 0.217799 & 0.217278 & 0.217574 \\
%4 & Metabolic & \bf 0.451658 & 0.441743 & 0.447843 \\
4 & Blogs & 0.432473 & 0.432259 & 0.432406 \\
5 & Airports2010 & 0.275524 & 0.275445 & 0.275502 \\
\hline

\end{tabular}
\\[10pt]
\end{center}
\end{table*}

%\end{article}
\end{document}